\documentclass[prd,twocolumn,twoside,preprintnumbers,superscriptaddress,nofootinbib,natbib,bm,amsrefs]{revtex4-2}
\usepackage{graphicx}
\usepackage{url}

\usepackage{xcolor}
\usepackage{amsmath, amstext, amsfonts,amssymb, amsthm}
\usepackage{comment}
\usepackage[utf8]{inputenc}
\usepackage{multirow}
\newcommand{\ov}{\overline}

\newcommand{\eg}{{\em e.g.}}
\newcommand{\ie}{{\em i.e.}}

\newcommand{\SM}{{\rm SM}}
\newcommand{\FB}{{\rm FB}}

\newcommand{\costl}{c_{\theta_l}}
\newcommand{\costa}{c_{\theta_\tau}}
\newcommand{\bottomed}{bottomed}

\newcommand{\bottomedIntro}{beauty}
\newcommand{\charmedIntro}{charming}

\definecolor{BlueViolet}{rgb}{0.2, 0.00, 0.7}
\definecolor{Blue}{rgb}{0.15, 0.00, 0.9}
\definecolor{lightblue}{rgb}{0.15, 0.35, 0.95}
\definecolor{kitgreen}{rgb}{0,
0.58823 
, 0.50980 
}
\usepackage[
colorlinks=true, linkcolor=lightblue,citecolor=lightblue,urlcolor=kitgreen]{hyperref}

\newcommand{\Eprint}[1]{\href{#1}}

\definecolor{lb}{rgb}{.74,.83,.9}
\definecolor{ly}{rgb}{1,.92,.8}
\definecolor{lr}{rgb}{.98,.85,.87}

\begin{document}
\preprint{KEK--TH--2729, P3H--25--038, TTP25--018}
\title{$b \to c$ semileptonic sum rule: Extension to angular observables}

\author{Motoi Endo}
\affiliation{KEK Theory Center, IPNS, KEK, Tsukuba 305--0801, Japan}
\affiliation{Graduate Institute for Advanced Studies, SOKENDAI, Tsukuba, Ibaraki 305--0801, Japan}
\affiliation{Kobayashi-Maskawa Institute for the Origin of Particles and the Universe, Nagoya University, Nagoya 464--8602, Japan}

\author{Syuhei Iguro}
\affiliation{Institute for Advanced Research, Nagoya University, Nagoya 464-8601, Japan}
\affiliation{Kobayashi-Maskawa Institute for the Origin of Particles and the Universe, Nagoya University, Nagoya 464--8602, Japan}

\author{Tim Kretz}
\affiliation{Institute for Theoretical Particle Physics (TTP), Karlsruhe Institute of Technology (KIT), Wolfgang-Gaede-Str.\,1, 76131 Karlsruhe, Germany}

\author{Satoshi Mishima}
\affiliation{Department of Liberal Arts, Saitama Medical University, Moroyama, Saitama 350-0495, Japan}

\author{Ryoutaro Watanabe}
\affiliation{Institute of Particle Physics and Key Laboratory of Quark and Lepton Physics (MOE), Central China Normal University, Wuhan, Hubei 430079, China}

\begin{abstract}
Lepton flavor universality is a key prediction of the Standard Model of particle physics and any violation of it immediately indicates the existence of new physics.
Given the recent more than $4\sigma$ discrepancy in charged current semileptonic $B$ meson decays and the absence of evident signals at the large hadron collider, independent cross-checks become invaluable.
In this context, $b \to c$ semileptonic sum rules based on heavy quark symmetry are interesting since they allow us to check the consistency of experimental results.
In this paper, we report newly found sum rules among angular observables of mesonic and baryonic $b\to c l \overline{\nu}$ decays holding exactly in the large mass limit of heavy quarks.
Moreover, we investigate corrections to the sum rule in realistic situations and discuss phenomenological implications. 
\\ 
---------------------------------------------------------------------------------------------------------------------------------
\end{abstract}
\maketitle

\section{Introduction}
\label{sec:intro}
The $b \to c$ semileptonic sum rule for $R_{H_c}={\rm{BR}}(H_b\to H_c \tau\ov\nu)/{\rm{BR}}(H_b\to H_c \ell\ov\nu)$ with $\ell = e,\mu$, is shown as \cite{Blanke:2018yud,Blanke:2019qrx,Fedele:2022iib,Duan:2024ayo}
\begin{align}
\frac{R_{\Lambda_c}}{R_{\Lambda_c}^{SM}}-\alpha_R \frac{R_{D}}{R_{D}^{SM}}-\beta_R \frac{R_{D^*}}{R_{D^*}^{SM}}=\delta_R,   
\label{eq:RSR}
\end{align}
where the coefficients satisfy $\alpha_R+\beta_R=1$ and are independent of new physics (NP) contributions. 
$\delta_R$ is found to be small compared to current experimental uncertainties, 
enabling us to apply it as a robust consistency check of experimental results.
Note that $B$ ($D$) and $B^*$ ($D^*$) form the lowest-lying \bottomedIntro\, (\charmedIntro) meson heavy quark doublet and $\Lambda_b$ ($\Lambda_c$) corresponds to the lowest-lying \bottomedIntro\, (\charmedIntro) baryon.
Hence, the sum rule is satisfied among ground-state to ground-state transitions.
Recently, based on the heavy quark effective theory (HQET) \cite{Isgur:1989vq,Neubert:1993mb}, another relation has been derived for mesonic and baryonic differential decay rates \cite{Endo:2025fke}
\begin{align}
    \frac{\kappa_{\Lambda_c}^{w}}{\zeta(w)^2}-\frac{2\left(\kappa_{D}^{w}+\kappa_{D^*}^{w}\right)}{(1+w)\xi(w)^2} \approx 0,
    \label{eq:DDRSR_simp}
\end{align}
where $\kappa_{H_c}^{w}=d\Gamma^{H_c}/dw$ with $\Gamma^{H_c}=\Gamma(H_b\to H_c \tau\ov\nu)$ are defined for $w = (m_{H_b}^2+m_{H_c}^2-q^2)/(2m_{H_b}m_{H_c})$ and $q^2$ being the invariant mass of the leptons.
Equation (\ref{eq:DDRSR_simp}) holds exactly in the large mass limit of heavy quarks, \ie, $m_{b,c} \gg \Lambda_{\rm QCD}$.
Also, Eq.\,(\ref{eq:DDRSR_simp}) can be rewritten as 
\begin{align}
    \frac{\kappa_{\Lambda_c}^{w}}{\kappa_{\Lambda_c}^{w,\SM}}-\alpha \frac{\kappa_{D}^{w}}{\kappa_{D}^{w,\SM}}-\beta\frac{\kappa_{D^*}^{w}}{\kappa_{D^*}^{w,\SM}} \approx 0,
    \label{eq:DDRSR}
\end{align}
again satisfying $\alpha +\beta=1$. 
Then, Eq.\,(\ref{eq:RSR}) is obtained by integrating numerators and denominators over $w$, and by normalizing with the decay widths of light-lepton modes.
In other words, the HQET is explicitly shown to be a pillar of the robust $b \to c$ semileptonic sum rule for $R_{H_c}$.

Given the more than $4\sigma$ discrepancy between the Standard Model (SM) prediction and the experimental value of $R_{D}$ and $R_{D^*}$ \cite{Crivellin:2025qsq}, a tremendous number of NP interpretations have been explored, and the relations which enable us to cross-check the consistency of the experimental results become more important.
In this paper, we provide sum rules involving angular observables, \eg, the forward-backward asymmetry of the charged lepton, 
\begin{align}
A_{\FB}^{H_cl }=\left(\Gamma^{H_c}_{c_{\theta_l}>0}-\Gamma^{H_c}_{c_{\theta_l}<0}\right)\biggl/\left(\Gamma^{H_c}_{c_{\theta_l}>0}+\Gamma^{H_c}_{c_{\theta_l}<0}\right),
\label{eq:AFB}
\end{align}
where $\Gamma^{H_c}_{c_{\theta_l} \gtrless 0}$ denotes the decay rates satisfying $\cos{\theta_l} \gtrless 0$ with $c_{\theta_l}$ being short for $\cos{\theta_l}$ and $\theta_l$ being the angle between the \bottomed-hadron and the charged lepton in the $W$-boson rest frame, 
as well as the charged-lepton polarization observable, 
\begin{align}
P_l^{H_c}=\left(\Gamma^{H_c}_{\lambda_l=1/2}-\Gamma^{H_c}_{\lambda_l=-1/2}\right)\biggl/\left(\Gamma^{H_c}_{\lambda_l=1/2}+\Gamma^{H_c}_{\lambda_l=-1/2}\right),
\label{eq:Pol}
\end{align}
where $\Gamma^{H_c}_{\lambda_l}$ is the rate with $\lambda_l$ denoting a charged-lepton polarization.
In the case of $l=\tau$, only the experimental measurement of $P_\tau^{D^*}$ is currently available and its uncertainty is still as large as ${\mathcal{O}}(100)\%$ \cite{Hirose:2017dxl}.
In future Belle II could determine $P_\tau^D$ at $3\%$ level and $P_\tau^{D^*}$ at about $15\%$ \cite{Alonso:2017ktd,Belle-II:2018jsg,Adamczyk:2019wyt}.
Baryon observables measured at on-going and future colliders \cite{LHCb:2018roe,Cerri:2018ypt,Ai:2024nmn,FCC:2025lpp} can be compared with the predictions via sum rules with up-coming Belle II input \cite{Belle-II:2018jsg}.

The outline of this letter is as follows:
In Sec.\,\ref{sec:SR}, we introduce our framework and present the new sum rule involving double-differential decay rates.
In Sec.\,\ref{sec:pheno}, we investigate corrections to the sum rule and discuss phenomenological implications.
Sec.\,\ref{sec:conc} is devoted to the conclusion.

\section{$b \to c$ semileptonic sum rule of angular observables}
\label{sec:SR}

We assume that the NP contributes to the $b\to c l\bar\nu_l$ transitions in the following form:
\begin{align}
 \label{eq:Hamiltonian}
 {\cal {L}}_{\rm{eff}}= -\frac{4  G_FV_{cb}}{\sqrt2}\biggl[ &(1+C_{V_L}^l)O_{V_L}^l+C_{S_L}^lO_{S_L}^l\\\nonumber
 &+C_{S_R}^lO_{S_R}^l+C_{T}^lO_{T}^l\biggl]\,.
\end{align}
The effective operators are defined at $\mu=\mu_b$ as
\begin{align}
 \label{eq:operator} 
 &O_{V_L}^l = (\overline{c} \gamma^\mu P_Lb)(\overline{l} \gamma_\mu P_L \nu_{l}),\,\,\, 
 O_{S_L}^l = (\overline{c}  P_Lb)(\overline{l} P_L \nu_{l}), \\[0.5em]
 &O_{S_R}^l = (\overline{c}  P_Rb)(\overline{l} P_L \nu_{l}),\,\,\, 
 O_{T}^l = (\overline{c}  \sigma^{\mu\nu}P_Lb)(\overline{l} \sigma_{\mu\nu} P_L \nu_{l}),\nonumber
\end{align}
where $l=e,\mu,\tau$, $P_{L(R)}=(1\mp\gamma_5)/2$, and $\sigma^{\mu\nu}=(i/2)[\gamma^\mu,\gamma^\nu]$ with $\sigma^{\mu\nu} \gamma_5 = - (i/2) \epsilon^{\mu\nu\rho\sigma} \sigma_{\rho\sigma}$.\footnote{
We assumed that the charged mediator of $b\to c l\ov\nu$ transition is much heavier than the typical energy scale of \bottomed-hadron decays.
Lepton flavor is assumed to be conserved. 
Also, all light neutrinos are assumed to be left-handed.
For studies including right-handed neutrinos see Refs.\,\cite{Iguro:2018qzf,Robinson:2018gza,Babu:2018vrl,Mandal:2020htr,Penalva:2021wye,Datta:2022czw,Bernlochner:2024xiz}.
} 
NP contributions are encoded into the Wilson coefficients (WCs) $C_X^l$ with $X=V_L$, $S_{L,R}$ and $T$.
The SM case corresponds to $C_{X}^l = 0$.
In the following, motivated by the discrepancy in $R_{D^{(*)}}$, we will focus on the tauonic NP scenario, \ie, $l=\tau$ \cite{Fedele:2022iib,Ray:2023xjn,Fedele:2023ewe}.

Let us provide a new sum rule holding among the double-differential decay rates, 
\begin{align}
    \frac{\kappa_{\Lambda_c}^{w\costl,\lambda_l}}{\zeta(w)^2}-\frac{2\left(\kappa_{D}^{w\costl,\lambda_l}+\kappa_{D^*}^{w\costl,\lambda_l}\right)}{(1+w)\xi(w)^2}=0,
    \label{eq:D3RSR}
\end{align}
where $\kappa_{H_c}^{w\costl,\lambda_l}=d^2\Gamma^{H_c}_{\lambda_l}/dw\,d\costl$\footnote{See Appendix\,\ref{sec:DDR} for details. 
Also, the sum rule holds for light-lepton modes as well.
} and the leading-order Isgur-Wise (IW) functions, $\zeta(w)$ and $\xi(w)$, are introduced for $\Lambda_b\to\Lambda_c$ and $B\to D^{(*)}$, respectively.
The relation holds for any $\costl$ and each $\lambda_l$ as well as any $w$ under the heavy quark symmetry, \ie, at the leading order of the IW expansion and the limit of heavy quark hadron masses satisfying $m_b \approx m_B \approx m_{\Lambda_b}$ and $m_c \approx m_D \approx m_{D^*} \approx m_{\Lambda_c}$.
Summing up both lepton polarizations and integrating over $\costl$ reproduces the single-differential decay rate sum rule, Eq.\,(\ref{eq:DDRSR_simp}).
This finding leads us to propose two new sum rules involving $A_{\FB}^\tau$ and $P_\tau$ for each to be tested experimentally:\footnote{Uncertainties from form factor as well as experimental measurements will be reduced by normalizing with light-lepton decay widths.}
\begin{itemize}
    \item{Sum rule for the forward-backward asymmetry of the charged lepton: $(A_\FB^{H_c\tau}/A_\FB^{H_c\tau,\,\rm{SM}})(R_{H_c}/R_{H_c}^{\rm{SM}})$.}
    \item{Sum rule for the $\tau$-polarization difference: $(P_{\tau}^{H_c}/P_{\tau}^{H_c,\rm{SM}})(R_{H_c}/R_{H_c}^{\rm{SM}})$.}
\end{itemize}
Their explicit forms are shown in Eqs.\,~(\ref{eq:PtaSR}) and (\ref{eq:AFBSR}) including corrections.
See Appendix\,\ref{sec:alpha} for the construction in the heavy quark limit. 
It is noted that the tauonic total decay width in the denominator of Eqs.\,(\ref{eq:AFB}) and (\ref{eq:Pol}) is canceled with that in the numerator of $R_{H_c}$.
$P_\tau^D$ is interesting since it is known to be useful to distinguish NP models which explain the $R_{D^{(*)}}$ anomaly \cite{Tanaka:1994ay,Nierste:2008qe,Tanaka:2010se,Sakaki:2013bfa,Iguro:2018vqb,Blanke:2018yud}.
Therefore confirming experimental results with the one involving $P^{H_c}_\tau$ provides an important cross-check.

\section{Corrections to the sum rule and implication}
\label{sec:pheno}

Let us check how large corrections exist in the sum rules.
In reality, the heavy quark symmetry is violated by higher-order corrections in the heavy quark expansion.
The heavy flavored hadron masses include corrections as well as the heavy quark mass.
In the HQET, they are expressed as \cite{Falk:1992wt,Falk:1992ws,Bernlochner:2018bfn}
\begin{align}
 m_{H_Q} \simeq m_Q \left(1+ \frac{\bar \Lambda}{m_Q} + \frac{\Delta m^2}{2m_Q^2} \right) \,,
 \label{eq:hadron_masses}
\end{align}
where $m_Q$ is a heavy quark mass parameter.
Also, $\bar \Lambda$ and $\Delta m$ parametrize a light quark contribution and the heavy quark kinetic energy in a hadron, respectively, and are of the order of the QCD scale $\Lambda_{\rm QCD}$.
Moreover, higher-order corrections to the IW functions of $\mathcal{O}(\Lambda_{\rm QCD}/2m_Q)$ and $\mathcal{O}(\Lambda_{\rm QCD}^2/4m_Q^2)$ enter in the $H_b\to H_c$ transition form factors.
We will use the HQET form factor input from Refs.\,\cite{Bernlochner:2018bfn,Iguro:2020cpg}.\footnote{To be precise we used the fit result of the z210 scenario for the $B\to D^{(*)}$ transition and the $\hat b_{1,2}\neq0$ scenario for the $\Lambda_b\to\Lambda_c$ transition. }
Then numerical formulae of $(P_{\tau}^{H_c}/P_{\tau}^{H_c,\rm{SM}})(R_{H_c}/R_{H_c}^{\rm{SM}})$ and $(A_\FB^{H_c\tau}/A_\FB^{H_c\tau,\,\rm{SM}})(R_{H_c}/R_{H_c}^{\rm{SM}})$ are obtained for mesonic and baryonic decays (see Eqs.\,(\ref{Eq:Generic_first})--(\ref{Eq:Generic_last}) in Appendix \ref{sec:generic}).
Combining three relations where the coefficients $\alpha$ and $\beta$ are determined such that $|1+C_{V_L}^\tau|^2$ and Re$[(1+C_{V_L}^\tau)C_{S_R}^{\tau*}]$ terms are vanishing (see, \eg, Refs.\,\cite{Blanke:2018yud,Blanke:2019qrx,Fedele:2022iib,Duan:2024ayo} for such a prescription), the sum rules are derived as
\begin{widetext}
\begin{align}
\label{eq:PtaSR}
&\frac{P_\tau^{\Lambda_c}}{P_\tau^{\Lambda_c,\SM}}\frac{R_{\Lambda_c}}{R_{\Lambda_c}^{\SM}}-\alpha_{P_\tau} \frac{P_\tau^{D}}{P_\tau^{D,\SM}}\frac{R_{D}}{R_{D}^{\SM}}-\beta_{P_\tau} \frac{P_\tau^{D^*}}{P_\tau^{D^*,\SM}}\frac{R_{D^*}}{R_{D^*}^{\SM}}=\delta_{P_\tau},  \\[0.5em]
\label{eq:AFBSR}
&\frac{A_\FB^{\Lambda_c\tau}}{A_\FB^{\Lambda_c\tau,\SM}}\frac{R_{\Lambda_c}}{R_{\Lambda_c}^{\SM}}-\alpha_{A_\FB^\tau} \frac{A_\FB^{D\tau}}{A_\FB^{D\tau,\SM}}\frac{R_{D}}{R_{D}^{\SM}}-\beta_{A_\FB^\tau} \frac{A_\FB^{D^*\tau}}{A_\FB^{D^*\tau,\SM}}\frac{R_{D^*}}{R_{D^*}^{\SM}}=\delta_{A_\FB^\tau}. 
\end{align}
If the corrections, $\delta_{P_\tau}$ and $\delta_{A_\FB^\tau}$, are small enough (even when their uncertainties are taken into account) compared to the experimental uncertainties, the sum rules are useful to check the consistency of the experimental results simply by neglecting the correction terms.
Even when the experimental values are determined so precisely that $\delta_{P_\tau}$ and $\delta_{A_\FB^\tau}$ are not negligible, the sum rules could be applied to discriminate the NP scenarios.
The corrections, $\delta_{P_\tau}$ and $\delta_{A_\FB^\tau}$, consist of bilinears of the WCs as 
$\delta_Y={\displaystyle \sum_{ij}} C_i^\tau C_j^{\tau *} \delta_Y^{ij}$
with $Y=P_\tau,\,A_\FB^\tau$, and are approximately expressed, ignoring uncertainties, as
\begin{align}
\label{eq:breakingPta}
    &\delta_{P_\tau}\simeq
    -0.149\textrm{Re}[(1+C_{V_L}^\tau )C_{S_L}^{\tau *}]
    -0.036\left(|C_{S_L}^\tau |^2+|C_{S_R}^\tau |^2\right)
    -0.136\textrm{Re}[C_{S_L}^\tau C_{S_R}^{\tau *}]
    +0.540\textrm{Re}[(1+C_{V_L}^\tau)C_T^{\tau *}]
    -0.892|C_T^{\tau}|^2,\\[0.5em]
\label{eq:breakingAFB}
    &\delta_{A_\FB^\tau}\simeq
    +2.73\textrm{Re}[(1+C_{V_L}^\tau )C_{S_L}^{\tau *}]
    +21.2\textrm{Re}[(1+C_{V_L}^\tau )C_T^{\tau *}]
    -46.2|C_T^\tau|^2
    -0.503\textrm{Re}[C_{S_L}^\tau C_T^{\tau *}]
    +11.2\textrm{Re}[C_{S_R}^\tau C_T^{\tau *}],
\end{align}
\end{widetext}
with $\alpha_{P_{\tau}}=1-\beta_{P_{\tau}}\simeq -0.258$, $\alpha_{A_{\FB}^\tau}=1-\beta_{A_{\FB}^\tau}\simeq 2.53$.
In the absence of the NP contribution, \ie, within the SM, the corrections vanish as $\delta_{P_\tau}=\delta_{A_\FB^\tau}=0$. 
Also, we obtain $P_\tau^{H_c,\SM} R_{H_c}^{\SM}\simeq 0.10, -0.12, -0.10$ and $A_\FB^{H_c\tau,\SM} R_{H_c}^{\SM}\simeq 0.11,\, -0.0094,\,-0.0082$ for $H_c=D,\,D^*,\,\Lambda_c$, respectively. 
In Appendix\,\ref{sec:plot}, Figs.\,\ref{fig:Pta} and \ref{fig:AFB} respectively show the probability distributions of $\alpha_Y$ and $\delta^{ij}_Y$ stemming from the form factor inputs.

For the central values, the coefficients involving the tensor operator in $\delta_{P_\tau}^{ij}$ are larger than the others, \eg, $\delta_{P_\tau}^{TT} \simeq -0.9$, while the scalar coefficients are smaller.
Besides, the tensor coefficients are $\mathcal{O}(10)$ in $\delta_{A_\FB^\tau}$, while $\delta_{A_\FB^\tau}^{V_LS_L}$ is about a few.
The former mainly comes from mismatches between $(A_\FB^{D^*\tau}/A_\FB^{D^*\tau,\SM})$ and $(A_\FB^{\Lambda_c\tau}/A_\FB^{\Lambda_c\tau,\SM})$ as seen from Eqs.\,(\ref{Eq:Generic_2ndlast}) and (\ref{Eq:Generic_last}).
In these numerical formulae, the coefficients are large because $A_\FB^{H_c\tau,\SM}$ is suppressed for $H_c = D^*,\,\Lambda_c$. 

As seen in Figs.~\ref{fig:Pta} and \ref{fig:AFB}, the uncertainty of $\alpha_{P_\tau}$ is less than $3\%$, while $\alpha_{A_\FB^\tau}$ has $\sim10\%$ uncertainty.
Regarding $\delta_{P_\tau}$, the uncertainties of the scalar coefficients are 30--$40\%$, while those of the tensor are $\sim 10\%$ for $ij=V_LT$ and $\sim 4\%$ for $TT$.
Similarly, in $\delta_{A_{\FB}^\tau}$, the uncertainty is $\sim 20\%$ except for the $ij=S_LT$ case where we have $\sim 100\%$ uncertainty.\footnote{In the $S_LT$ case, the relative uncertainty is large, because the central value is suppressed by a cancellation among the contributions.}
\begin{table}[b]
\vspace{.2cm}
\begin{center}
\scalebox{0.975}{
  \begin{tabular}{cccccc} 
 Scenario & Parameter & Value &  Pull & $\delta_{P_\tau}$ &  $\delta_{A_\FB^\tau}$  \\ \hline
$S_R$ & $C_{S_R}^\tau$ & $0.18$ & $3.9$ & $-0.001$ & $0$\\
$S_L$ & $C_{S_L}^\tau$ & $-0.57\pm0.86i$ & $4.3$ & $0.05$ & $-1.6$\\
$T$ & $C_{T}^\tau$ & $0.02 \pm 0.13i$ & $3.8$  & $-0.004$&$-0.38$\\
${\rm{R}}_2$ & $C_{S_L}^\tau=8.4C_T^\tau$ & $-0.09 \pm 0.56i$ & $4.4$  &$-0.008$ & $-0.7$\\
${\rm{S}}_1$ & $C_{S_L}^\tau=-8.9C_T^\tau$ & $0.18$ & $4.1$  &$-0.04$ &$0.05$\\
${\rm{U}}_1$ & $C_{V_L}^\tau$,\,$\phi$ & $0.075,\,\pm 0.466\pi$ & $4.4$  &$-0.003$& 0\\ \hline
\end{tabular}
}
  \caption{
  \vspace{.1cm}
  Corrections to the sum rules, $\delta_{P_\tau}$ and $\delta_{A_\FB^\tau}$, in the single operator ($S_R, S_L, T$) and single leptoquark scenarios (${\rm{R}}_2, {\rm{S}}_1, {\rm{U}}_1$).
  The first column represents the scenario, whose relevant WCs are listed in the second. 
  For ${\rm{U}}_1$ LQ, we consider the $U(2)$-flavored scenario, satisfying $C_{S_R}^\tau=-3.7e^{i\phi}C_{V_L}^\tau$. 
  See Ref.\,\cite{Iguro:2024hyk} for the detail.
  The best-fit values of the WCs at the $\mu_b$ scale are shown in the third, and the fit quality is expressed by the pull in the forth, whose definition is found in Ref.\,\cite{Iguro:2024hyk}. 
  The last two columns provide the central values of $\delta_{P_\tau}$ and $\delta_{A_\FB^\tau}$ in each scenario. 
 }
  \label{Tab:result}
\end{center}   
\end{table}

Let us estimate $\delta_{P_\tau}$ and $\delta_{A_\FB^\tau}$ with the WCs which are determined by performing a global fit to the current experimental values of $R_{D^{(*)}}$ and $F_L^{D^*}$ \cite{Iguro:2024hyk}.
The results are summarized in Table \ref{Tab:result}.
We consider three `single operator' scenarios and three `single leptoquark (LQ)' scenarios. 
In Appendix\,\ref{sec:plot}, Figs.\,\ref{fig:delPtaNP} and \ref{fig:delAFBNP} respectively show $\delta_{P_\tau}$ and $\delta_{A_\FB^\tau}$ at each benchmark point.
We define the central value by fitting the probabilities to Gaussian distributions. 
Out of three single operator scenarios, the $S_{L}$ case predicts $\delta_{P_\tau}\sim 0.05$, while the others predict $\delta_{P_\tau} = {\mathcal{O}}(10^{-3})$.
Among three LQ scenarios, only the ${\rm{S}}_1$ LQ scenario yields $\delta_{P_\tau}\sim -0.04$, while the other two scenarios predict at most $1\%$ level in $\delta_{P_\tau}$.
On the other hand, $\delta_{A_{\FB}^\tau}$ can be largely deviated from 0 and become as large as ${\mathcal{O}}(-1)$ in the $S_{L}$ and ${\rm{R}}_2$ LQ scenarios.
The $T$ and ${\rm{S}}_1$ LQ scenarios predict smaller corrections.
It is noted that the $S_R$ and ${\rm{U}}_1$ LQ scenarios predict $\delta_{A_{\FB}^\tau}=0$ and hence they are removed from the figure.

In summary, the correction to the sum rule $\delta_Y$ can be large for $Y=A_\FB^\tau$ and at most $\sim 5\%$ for $Y=P_\tau$.
This means that the sum rule involving $A_\FB^\tau$ is more sensitive to the NP effect and can be useful to discriminate the scenarios if the experimental uncertainties are comparable.
While checking the experimental consistency with the $\tau$-polarization sum rule will be important too.
Currently, future experimental prospects are available only for $P_\tau^D$ and $P_\tau^{D^*}$.
Although estimating the uncertainty of the product of observables, \eg, $P_\tau^{H_c}R_{H_c}$, needs careful study as some of the uncertainties in each observable should correlate to each other, it is highly encouraged in light of the new angular sum rules.

\section{Conclusion}
\label{sec:conc}

In this paper, we extended the previous studies of the $b\to c$ semileptonic sum rule for the single-differential decay rates and found a new relation for the double-differential decay rates, which holds exactly in the heavy quark limit.
We then derived two sum rules involving the $\tau$-polarization observable, $P_\tau$, and the charged lepton forward-backward asymmetry, $A_\FB^\tau$.
In reality, the heavy quark symmetry is violated and the sum rules receive corrections from higher-order effects in the heavy quark expansion, \ie, the realistic mass spectrum and the inclusion of higher-order terms of the IW functions.
We also evaluated these corrections numerically.
Using HQET-based form factors, we demonstrated that currently the sum rule coefficients, $\alpha_{P_\tau}$ and $\alpha_{A_\FB^\tau}$, can be determined with precisions of approximately $3\%$ and $11\%$, respectively.
To reduce the uncertainty of these coefficients, experimental input of $\Lambda_b\to \Lambda_c \mu\ov\nu$ and Lattice calculations are important.
Furthermore, we estimated the corrections in the NP scenarios motivated by the $R_{D^{(*)}}$ anomaly.
It is found that these corrections are at most $\sim 5\%$ for the $\tau$-polarization sum rule, Eq.\,(\ref{eq:PtaSR}), while they can reach around $-100\%$ for the $A_\FB^\tau$ case, Eq.\,(\ref{eq:AFBSR}), depending on the NP scenarios.
These new sum rules will provide independent cross-checks of experimental results.
By testing both angular sum rules, we can better explore and tell apart NP scenarios.
These results encourage the feasibility study at on-going and future experiments \cite{Belle-II:2018jsg,LHCb:2018roe,Cerri:2018ypt,Ai:2024nmn,FCC:2025lpp}.

\section*{Acknowledgements}
The authors thank Ulrich Nierste, Wen-Feng Duan, Martin S. Lang, Teppei Kitahara, Hiroyasu Yonaha, Florian Kretz, Andreas Crivellin, and Takashi Kaneko for the inspiring discussions and for encouraging this project.
This work is supported by JSPS KAKENHI Grant Numbers 22K21347 [M.E. and S.I.], 24K07025 [S.M.], 24K22879 [S.I.], 24K23939 [S.I.] and 25K17385 [S.I.], and by the Deutsche Forschungsgemeinschaft (DFG, German Research Foundation) under grant 396021762 - TRR 257 [T.K.]. 
The work is also supported by JPJSCCA20200002 and the Toyoaki scholarship foundation [S.I.].
We also appreciate KEK-KMI joint appointment program [M.E. and S.I.] and KMI grant for a young researcher [S.I.] which accelerated this project. 

\begin{widetext}
\appendix
\section{Differential decay rate}
\label{sec:DDR}

Here, let us write the double differential decay rates of $B\to D^{(*)} l\bar\nu$ and $\Lambda_b\to\Lambda_cl\bar\nu$ for each $l$-lepton polarization such that
\begin{align}
 {d^2\Gamma^{\lambda_l} (B\to D l\bar\nu) \over dq^2d\cos\theta_l} 
 & =  {G_F^2 |V_{cb}|^2\eta_{\rm EW} \sqrt{Q_+^DQ_-^D} \over 256 \pi^3 m_{B}^3} q^2 \left(1 - {m_l^2 \over q^2} \right)^2 \left( \mathcal A_0^{\lambda_l} + \mathcal A_1^{\lambda_l} \cos\theta_l + \mathcal A_2^{\lambda_l} \cos^2\theta_l \right) \,, \\[0.5em]
  {d^2\Gamma^{\lambda_l} (B\to D^{*} l\bar\nu) \over dq^2d\cos\theta_l} 
 & =  {G_F^2 |V_{cb}|^2 \eta_{\rm EW}\sqrt{Q_+^{D^*}Q_-^{D^*}} \over 512 \pi^3 m_{B}^3} q^2 \left(1 - {m_l^2 \over q^2} \right)^2 \left( \mathcal B_0^{\lambda_l} + \mathcal B_1^{\lambda_l} \cos\theta_l + \mathcal B_2^{\lambda_l} \cos^2\theta_l \right) \,, \\[0.5em]
 {d^2\Gamma^{\lambda_l} (\Lambda_b\to\Lambda_cl\bar\nu) \over dq^2d\cos\theta_l} 
 & = {G_F^2 |V_{cb}|^2 \eta_{\rm EW}\sqrt{Q_+^{\Lambda_c}Q_-^{\Lambda_c}} \over 512 \pi^3 m_{\Lambda_b}^3} q^2 \left(1 - {m_l^2 \over q^2} \right)^2 \left( \mathcal C_0^{\lambda_l} + \mathcal C_1^{\lambda_l} \cos\theta_l + \mathcal C_2^{\lambda_l} \cos^2\theta_l \right) \,, 
\end{align}
with $\eta_{\rm EW}$ being an electroweak correction and 
\begin{align}
 Q_\pm^{H_c} =\left(m_{H_b} \pm m_{H_c}\right)^2 -q^2 \,. 
\end{align}
The squared-amplitudes for $B\to D l\bar\nu$ are given by
\begin{align}
 \mathcal A^{+1/2}_0
 =  
 &~ |1+C_{V_L}^l|^2 {m_l^2 \over q^2} (H^s_{V_t})^2
 + |C_{S_L}^l+C_{S_R}^l|^2 (H^s_S)^2 
 +2 \text{Re}[(1+C_{V_L}^l) (C_{S_L}^l+C_{S_R}^l)^*] {m_l \over \sqrt{q^2}} H^s_{V_t} H^s_S \,, \\[1em]
\mathcal A^{+1/2}_1
 =
 &~ 2 |1+C_{V_L}^l|^2 {m_l^2 \over q^2} H^s_{V_t} H^s_{V_0} 
 -8 \text{Re}[(1+C_{V_L}^l) C_{T}^{l*}] {m_l \over \sqrt{q^2}} H^s_{V_t} H^s_{T}  \\[0.5em]
 & +2 \text{Re}[(1+C_{V_L}^l) (C_{S_L}^l+C_{S_R}^l)^*] {m_l \over \sqrt{q^2}} H^s_{V_0} H^s_S
 -8 \text{Re}[ (C_{S_L}^l+C_{S_R}^l) C_T^{l*}] H^s_S H^s_{T} \,,\notag \\[1em]
\mathcal A^{+1/2}_2
 = 
 &~ |1+C_{V_L}^l|^2 {m_l^2 \over q^2} (H^s_{V_0})^2
 +16 |C_T^l|^2  (H^s_{T})^2
 -8 \text{Re}[(1+C_{V_L}^l) C_{T}^{l*}] {m_l \over \sqrt{q^2}} H^s_{V_0} H^s_{T} \,, \\[1em]
 \mathcal A^{-1/2}_0 = &~ - \mathcal A^{-1/2}_2
 =
 |1+C_{V_L}^l|^2 (H^s_{V_0})^2
 +16|C_T^l|^2 {m_l^2 \over q^2}  (H^s_{T})^2
 -8\text{Re}[(1+C_{V_L}^l) C_{T}^{l*}] {m_l \over \sqrt{q^2}} H^s_{V_0} H^s_{T} \,, \\[1em]
 \mathcal A^{-1/2}_1 = &~ 0\,.
\end{align}
For $B\to D^{*} l\bar\nu$, they are given by
\begin{align}
 \mathcal B^{+1/2}_0
 =  
 &~ |1+C_{V_L}^l|^2 {m_l^2 \over q^2} \big( (H_{V_+})^2 + (H_{V_-})^2 + 2(H_{V_t})^2 \big) 
 +2 |C_{S_L}^l-C_{S_R}^l|^2 H_S^2 
 + 16 |C_T^l|^2 \, \big( (H_{T_+})^2 + (H_{T_-})^2 \big) \notag \\[0.5em]
 & -4\text{Re}[(1+C_{V_L}^l) (C_{S_L}^l-C_{S_R}^l)^*] {m_l \over \sqrt{q^2}} H_{V_t} H_S
 -8\text{Re}[(1+C_{V_L}^l) C_{T}^{l*}] {m_l \over \sqrt{q^2}} \big( H_{V_+} H_{T_+} - H_{V_-} H_{T_-} \big) \,, \\[1em]
 \mathcal B^{+1/2}_1
 =
 &~4|1+C_{V_L}^l|^2 {m_l^2 \over q^2} H_{V_0} H_{V_t} 
 -4\text{Re}[(1+C_{V_L}^l) (C_{S_L}^l-C_{S_R}^l)^*] {m_l \over \sqrt{q^2}} H_{V_0} H_S \notag \\[0.5em]
 & -16 \text{Re}[(1+C_{V_L}^l) C_{T}^{l*}] {m_l \over \sqrt{q^2}} H_{V_t} H_{T_0}
 +16\text{Re}[ (C_{S_L}^l-C_{S_R}^l) C_T^{l*}] H_S H_{T_0} \,, \\[1em]
 \mathcal B^{+1/2}_2
 =
 &~ |1+C_{V_L}^l|^2 {m_l^2 \over q^2} \big( 2(H_{V_0})^2 - (H_{V_+})^2 - (H_{V_-})^2 \big) 
 + 16 |C_T^l|^2 \, \big( 2(H_{T_0})^2 - (H_{T_+})^2 - (H_{T_-})^2 \big) \notag \\[0.5em]
 & - 8\text{Re}[(1+C_{V_L}^l) C_{T}^{l*}] {m_l \over \sqrt{q^2}} \big( 2H_{V_0} H_{T_0} - H_{V_+} H_{T_+} + H_{V_-} H_{T_-} \big) \,, \\[1em]
 \mathcal B^{-1/2}_0
 =
 &~ |1+C_{V_L}^l|^2 \big( 2(H_{V_0})^2 + (H_{V_+})^2 + (H_{V_-})^2 \big) 
 + 16 |C_T^l|^2 \, {m_l^2 \over q^2} \big( 2(H_{T_0})^2 + (H_{T_+})^2 + (H_{T_-})^2 \big) \notag \\[0.5em]
 & - 8\text{Re}[(1+C_{V_L}^l) C_{T}^{l*}] {m_l \over \sqrt{q^2}} \big( 2H_{V_0} H_{T_0} + H_{V_+} H_{T_+} - H_{V_-} H_{T_-} \big) \,,  \\[1em]
 \mathcal B^{-1/2}_1
 =
 &~ 2 |1+C_{V_L}^l|^2 \big( (H_{V_+})^2 - (H_{V_-})^2 \big) 
 +32 |C_T^l|^2 \, {m_l^2 \over q^2} \big( (H_{T_+})^2 - (H_{T_-})^2 \big) \notag \\[0.5em] 
 & - 16\text{Re}[(1+C_{V_L}^l) C_{T}^{l*}] {m_l \over \sqrt{q^2}} \big( H_{V_+} H_{T_+} + H_{V_-} H_{T_-} \big) \,, \\[1em]
 \mathcal B^{-1/2}_2
 =
 &~ |1+C_{V_L}^l|^2 \big( -2(H_{V_0})^2 + (H_{V_+})^2 + (H_{V_-})^2 \big) 
 + 16 |C_T^l|^2 \, {m_l^2 \over q^2} \big( -2(H_{T_0})^2 + (H_{T_+})^2 + (H_{T_-})^2 \big) \notag \\[0.5em]
 & + 8\text{Re}[(1+C_{V_L}^l) C_{T}^{l*}] {m_l \over \sqrt{q^2}} \big( 2H_{V_0} H_{T_0} - H_{V_+} H_{T_+} + H_{V_-} H_{T_-} \big) \,.
\end{align}
 The squared-amplitudes for $\Lambda_b\to\Lambda_cl\bar\nu$ are given by
\begin{align}
 \mathcal C^{+1/2}_0
 =  
 &~ |1+C_{V_L}^l|^2 {m_l^2 \over q^2} \big( (H^{H+}_{V_\perp})^2 + (H^{H-}_{V_\perp})^2 + (H^{H+}_{V_0})^2 + (H^{H-}_{V_0})^2 \big)
 + 16|C_T^l|^2 \big( { (H^{H+}_{T_\perp})^2 + (H^{H-}_{T_\perp})^2} \big) \notag \\[0.5em]
 & + (|C_{S_L}^l|^2 + |C_{S_R}^l|^2) \big( (H^{H+}_{S})^2 + (H^{H-}_{S})^2 \big)
 + 4\text{Re}[C_{S_L}^l C_{S_R}^{l*}] H^{H+}_{S} H^{H-}_{S} \notag \\[0.5em] 
 & + 2\text{Re}[(1+C_{V_L}^l) C_{S_L}^{l*}] {m_l \over \sqrt{q^2}} \big( H^{H+}_{V_0} H^{H+}_{S} +H^{H-}_{V_0} H^{H-}_{S} \big) 
 + 2\text{Re}[(1+C_{V_L}^l) C_{S_R}^{l*}] {m_l \over \sqrt{q^2}} \big( H^{H+}_{V_0} H^{H-}_{S} +H^{H-}_{V_0} H^{H+}_{S} \big) \notag \\[0.5em]
 & + 8\text{Re}[(1+C_{V_L}^l) C_{T}^{l*}] {m_l \over \sqrt{q^2}} \big( H^{H+}_{V_\perp} H^{H+}_{T_\perp} + H^{H-}_{V_\perp} H^{H-}_{T_\perp} \big) \,, \\[1em]
 \mathcal C^{+1/2}_1
 =  
 &~ 2 |1+C_{V_L}^l|^2 {m_l^2 \over q^2} \big( H^{H+}_{V_0} H^{H+}_{V_+} + H^{H-}_{V_0} H^{H-}_{V_+} \big) 
 + 2\text{Re}[(1+C_{V_L}^l) C_{S_L}^{l*}] {m_l \over \sqrt{q^2}} \big( H^{H+}_{V_+} H^{H+}_{S} + H^{H-}_{V_+} H^{H-}_{S} \big) \notag \\[0.5em]
 & +2\text{Re}[(1+C_{V_L}^l) C_{S_R}^{l*}] {m_l \over \sqrt{q^2}} \big( H^{H+}_{V_+} H^{H-}_{S} + H^{H-}_{V_+} H^{H+}_{S} \big) 
 +8\text{Re}[(1+C_{V_L}^l) C_{T}^{l*}] {m_l \over \sqrt{q^2}} \big( H^{H+}_{V_0} H^{H+}_{T_+} + H^{H-}_{V_0} H^{H-}_{T_+} \big) \notag \\[0.5em] 
 & + 8\text{Re}[ C_{S_L}^l C_{T}^{l*}] \left( H^{H+}_{T_+} H^{H+}_{S}+H^{H-}_{T_+} H^{H-}_{S} \right) 
 + 8\text{Re}[ C_{S_R}^l C_{T}^{l*}]  \left(H^{H+}_{T_+} H^{H-}_{S}+H^{H-}_{T_+} H^{H+}_{S} \right)\,,  \\[1em]
 \mathcal C^{+1/2}_2
 = 
 & |1+C_{V_L}^l|^2 {m_l^2 \over q^2} \big( (H^{H+}_{V_+})^2 + (H^{H-}_{V_+})^2 - (H^{H+}_{V_\perp})^2 - (H^{H-}_{V_\perp})^2 \big) 
 +16 |C_T^l|^2 \big( (H^{H+}_{T_+})^2 + (H^{H-}_{T_+})^2 - (H^{H+}_{V_\perp})^2 - (H^{H-}_{V_\perp})^2 \big)\nonumber \\[0.5em]
 & + 8\text{Re}[(1+C_{V_L}^l) C_{T}^{l*}] {m_l \over \sqrt{q^2}} \big( H^{H+}_{V_+} H^{H+}_{T_+} + H^{H-}_{V_+} H^{H-}_{T_+} - H^{H+}_{V_\perp} H^{H+}_{T_\perp} - H^{H-}_{V_\perp} H^{H-}_{T_\perp} \big) \,,\\[0.5cm]
 \mathcal C^{-1/2}_0
 = 
 & |1+C_{V_L}^l|^2 \big( (H^{H+}_{V_+})^2 + (H^{H-}_{V_+})^2 + (H^{H+}_{V_\perp})^2 + (H^{H-}_{V_\perp})^2 \big)
 +16 |C_T^l|^2 {m_l^2 \over q^2} \big( (H^{H+}_{T_+})^2 + (H^{H-}_{T_+})^2 + (H^{H+}_{T_\perp})^2 + (H^{H-}_{T_\perp})^2 \big) \nonumber \\[0.5em]
 & + 8\text{Re}[(1+C_{V_L}^l) C_{T}^{l*}] {m_l \over \sqrt{q^2}} \big( H^{H+}_{V_+} H^{H+}_{T_+} + H^{H-}_{V_+} H^{H-}_{T_+} + H^{H+}_{V_\perp} H^{H+}_{T_\perp} + H^{H-}_{V_\perp} H^{H-}_{T_\perp} \big) \,,\\[0.5cm]
 \mathcal C^{-1/2}_1
 =  
 & 2|1+C_{V_L}^l|^2 \big( (H^{H+}_{V_\perp})^2 - (H^{H-}_{V_\perp})^2 \big)
 + 32|C_T^l|^2 {m_l^2 \over q^2} \big( (H^{H+}_{T_\perp})^2 - (H^{H-}_{T_\perp})^2 \big) \nonumber \\[0.5em]
 & + 16\text{Re}[(1+C_{V_L}^l) C_{T}^{l*}] {m_l \over \sqrt{q^2}} \big( H^{H+}_{V_\perp} H^{H+}_{T_\perp} - H^{H-}_{V_\perp} H^{H-}_{T_\perp} \big)\,, \\[0.5cm]
 \mathcal C^{-1/2}_2
 = 
 & |1+C_{V_L}^l|^2 \big( (H^{H+}_{V_\perp})^2 + (H^{H-}_{V_\perp})^2 - (H^{H+}_{V_+})^2  - (H^{H-}_{V_+})^2 \big)
 +16|C_T^l|^2 {m_l^2 \over q^2} \big( (H^{H+}_{T_\perp})^2 + (H^{H-}_{T_\perp})^2 - (H^{H+}_{T_+})^2 - (H^{H-}_{T_+})^2 \big)  \nonumber \\[0.5em]
 & +8\text{Re}[(1+C_{V_L}^l) C_{T}^{l*}] {m_l \over \sqrt{q^2}} \big( H^{H+}_{V_\perp} H^{H+}_{T_\perp} + H^{H-}_{V_\perp} H^{H-}_{T_\perp} - H^{H+}_{V_+} H^{H+}_{T_+} - H^{H-}_{V_+} H^{H-}_{T_+} \big) \,. 
\end{align} 
The hadronic helicity amplitudes, $H^s_X$ and $H_X$, for $B\to D^{(*)} l\ov\nu$ are given in the HQET description as\,\cite{Tanaka:2012nw,Sakaki:2013bfa}
\begin{align}
 H_{V_0}^s &= m_B \sqrt{\frac{r_D(w^2-1)}{\hat q_D^2}} \Big[ (1+r_D)h_+ -(1-r_D)h_- \Big] \,, \notag \\[0.5em]
 H_{V_t}^s &= m_B \sqrt{\frac{r_D}{\hat q_D^2}} \Big[ (1-r_D)(w+1)h_+ -(1+r_D)(w-1)h_- \Big] \,,  \\[0.5em]
 H_{S}^s &= m_B \sqrt{r_D} (w+1) h_S \,, \notag \\[0.5em]
 H_{T}^s &= - m_B \sqrt{r_D(w^2-1)}\, h_T\,,\notag
\end{align}
and
\begin{align}
 H_{V_\pm} 
 &= m_B \sqrt{r_{D^*}} \Big[ (w+1) h_{A_1} \mp \sqrt{w^2-1} h_{V} \Big] \,, 
 \notag \\[0.5em]
 H_{V_0}
 &= m_B \sqrt{\frac{r_{D^*}}{\hat q_{D^*}^2}} (w+1) 
 \Big[ (r_{D^*}-w) h_{A_1} + (w-1) (r_{D^*} h_{A_2} + h_{A_3}) \Big] \,, 
 \notag \\[0.5em]
 H_{V_t} 
 &= -m_B \sqrt{\frac{r_{D^*}(w^2-1)}{\hat q_{D^*}^2}}
 \Big[ (w+1) h_{A_1} + (r_{D^*}w-1) h_{A_2} + (r_{D^*}-w) h_{A_3} \Big] \,, \\[0.5em]
 H_{S} 
 &= - m_B \sqrt{r_{D^*}(w^2-1)} h_P \,, 
 \notag \\ 
 H_{T_\pm} 
 &= \pm m_B \sqrt{\frac{r_{D^*}}{\hat q_{D^*}^2}} \left[ 1-r_{D^*} (w \mp \sqrt{w^2-1}) \right]
 \left[ h_{T_1} + h_{T_2} + \left( w \pm \sqrt{w^2-1} \right) (h_{T_1} - h_{T_2}) \right] \,, 
 \notag \\[0.5em]
 H_{T_0}
 &= -m_B \sqrt{r_{D^*}} \Big[ (w+1) h_{T_1} +(w-1) h_{T_2} +2(w^2-1) h_{T_3} \Big] \notag  \,,
\end{align}
where $\hat q_{H_c}^2 = q^2/m_{H_b}^2 = 1 - 2r_{H_c}w + r_{H_c}^2$ and $r_{H_c} = m_{H_c}/m_{H_b}$ are defined.
Regarding $\Lambda_b\to\Lambda_c l\ov\nu$, they are summarized as~\cite{Datta:2017aue,Bernlochner:2018bfn}
\begin{align}
 H^{H\pm}_{V_+}  
 &= m_{\Lambda_b} \sqrt{\frac{2r_\Lambda}{\hat q_\Lambda^2}} 
 \Big\{
 \sqrt{w-1} \Big[(1+r_\Lambda)f_1 + (w+1)(f_2 r_\Lambda + f_3)\Big]
 \mp \sqrt{w+1} \Big[(1-r_\Lambda)g_1 - (w-1)(g_2 r_\Lambda + g_3)\Big]
 \Big\}\,,  \nonumber \\[0.5em]
 H^{H\pm}_{V_\perp} 
 &= m_{\Lambda_b} \sqrt{2r_\Lambda} \Big[ \sqrt{w-1} f_1 \mp \sqrt{w+1} g_1 \Big] \,,
  \\[0.5em]
 H^{H\pm}_{V_0}
 &= m_{\Lambda_b} \sqrt{\frac{2r_\Lambda}{\hat q_\Lambda^2}} \Big\{
 \sqrt{w+1} \Big[(1-r_\Lambda)f_1 + f_2(1-w r_\Lambda) + f_3(w-r_\Lambda)\Big]
 \mp \sqrt{w-1} \Big[(1+r_\Lambda)g_1 - g_2(1-w r_\Lambda) - g_3(w-r_\Lambda)\Big] \Big\},
 \nonumber \\[0.5em]
 H_{S}^{H\pm} 
 &= m_{\Lambda_b} \sqrt{2r_\Lambda} \left(\sqrt{w+1}h_S^\prime\pm \sqrt{w-1}h_P^\prime\right)  \,, 
 \nonumber \\[0.5em]
 H^{H\pm}_{T_+}
 &=  m_{\Lambda_b} \sqrt{2r_\Lambda} \Big\{ \sqrt{w-1} \Big[h_1-h_2+h_3-(w+1)h_4\Big] \pm \sqrt{w+1}h_1 \Big\}\, ,
 \nonumber \\[0.5em]
 H^{H\pm}_{T_\perp}
 &= m_{\Lambda_b} \sqrt{\frac{2r_\Lambda}{\hat q_\Lambda^2}} 
 \Big\{ 
 \sqrt{w-1} \Big[ (1+r_\Lambda) h_1 - (1-wr_\Lambda) h_2 - (w-r_\Lambda) h_3 \Big] 
 \pm \sqrt{w+1} \Big[ (1-r_\Lambda) h_1 - (w-1) (h_2 r_\Lambda + h_3) \Big]  \Big\} 
 \,.\nonumber
\end{align}
To be self-complete, we introduce the HQET form factors as 
\begin{align}
 \langle D | \bar c \gamma^\mu b | B \rangle 
 & = \sqrt{m_B m_D} \big[ h_+ (v+v')^\mu + h_- (v-v')^\mu \big] \,, 
 \notag \\[0.5em]
 \langle D |\bar c b| B \rangle
 & = \sqrt{m_B m_D} (w+1) h_S \,, 
 \\[0.5em]
 \langle D |\bar c \gamma^\mu \gamma_5 b| B \rangle
 & = \langle D |\bar c \gamma_5 b| B \rangle = 0 \,, 
 \notag \\[0.5em]
 \langle D |\bar c \sigma^{\mu\nu} b| B \rangle
 & = -i \sqrt{m_B m_D}\, h_T \big( v^\mu v^{\prime\nu} - v^{\prime\mu} v^\nu  \big) \,, 
 \notag
\end{align} 
for the $B\to D$ transitions, and 
\begin{align}
 \langle D^* | \bar c \gamma^\mu b | B \rangle
 & = i \sqrt{m_B m_{D^*}} h_V \varepsilon^{\mu\nu\rho\sigma} \epsilon^*_\nu v'_\rho v_\sigma \,, 
  \notag \\[0.5em]
 \langle D^* | \bar c \gamma^\mu \gamma_5 b | B \rangle 
 & = \sqrt{m_B m_{D^*}} \big[ h_{A_1} (w+1) \epsilon^{*\mu} - (\epsilon^* \cdot v) \left( h_{A_2} v^\mu + h_{A_3} v^{\prime\mu} \right) \big] \,, 
   \notag\\[0.5em]
  \langle D^* | \bar c \gamma_5 b | B \rangle 
 & = -\sqrt{m_B m_{D^*}} (\epsilon^* \cdot v) h_P \,, 
 \notag \\[0.5em]
 \langle D^* |\bar c b| B \rangle
 & = 0 \,, 
  \\[0.5em]
 \langle D^* | \bar c \sigma^{\mu\nu} b | B \rangle 
 & = -\sqrt{m_B m_{D^*}} \varepsilon^{\mu\nu\rho\sigma} 
 \big[ h_{T_1} \epsilon^{*}_\rho (v+v')_\sigma + h_{T_2} \epsilon^{*}_\rho (v-v')_\sigma 
 + h_{T_3} (\epsilon^* \cdot v) (v+v')_\rho (v-v')_\sigma \big] \notag\,,
\end{align} 
for $B\to D^{*}$.
For $\Lambda_b\to \Lambda_b l\ov\nu$, they are given as
\begin{align}
 \langle \Lambda_c| \bar c\gamma_\mu b |\Lambda_b\rangle 
 &= \bar u(p',s') \big[ f_1 \gamma_\mu + f_2 v_\mu + f_3 v'_\mu \big] u(p,s) \,, 
 \notag \\[0.5em]
 \langle \Lambda_c| \bar c\gamma_\mu\gamma_5 b |\Lambda_b\rangle 
 &= \bar u(p',s') \big[ g_1 \gamma_\mu + g_2 v_\mu + g_3 v'_\mu \big] \gamma_5\, u(p,s) \,, 
 \notag \\[0.5em]
 \langle \Lambda_c| \bar c\, b |\Lambda_b\rangle 
 &= h_S^\prime\, \bar u(p',s')\, u(p,s)
 \,,  \\[0.5em]
 \langle \Lambda_c| \bar c \gamma_5 b |\Lambda_b\rangle 
 &= h_P^\prime\, \bar u(p',s')\, \gamma_5\, u(p,s) \,, 
 \notag \\[0.5em]
 \langle \Lambda_c| \bar c\, \sigma_{\mu\nu}\, b |\Lambda_b\rangle 
 &= \bar u(p',s') \big[ h_1\, \sigma_{\mu\nu}
 + i\, h_2 (v_\mu \gamma_\nu - v_\nu \gamma_\mu)
 + i\, h_3 (v'_\mu \gamma_\nu - v'_\nu \gamma_\mu)
 + i\, h_4 (v_\mu v'_\nu - v_\nu v'_\mu) \big] u(p,s) \,, \notag
\end{align}
where $u(p,s)$ are spinors with momentum $p$ and spin $s$.
Moreover, $v = p/m_{H_b}$ and $v' = p'/m_{H_c}$ satisfying $ v \cdot v' = w = (m_{H_b}^2 + m_{H_c}^2 - q^2)/(2m_{H_b} m_{H_c})$ are introduced.
The form factors $f_i$, $g_i$, and $h_i$ are functions of $w$ and expressed in the heavy quark limit as~\cite{Isgur:1989vq,Isgur:1990pm}
\begin{align}
 \label{eq:DFFinHQL}
 & h_+ = h_V = h_{A_1} = h_{A_3} = h_S = h_P = h_T = h_{T_1} = \xi(w) \,, \notag\\[0.5em]
 & h_- = h_{A_2} = h_{T_2} = h_{T_3} = 0 \,,\\[0.5em]
 & f_1 = g_1 = h_S^\prime = h_P^\prime = h_1 = \zeta(w) \,, \notag\\[0.5em]
 & f_2 = f_3 = g_2 = g_3 = h_2 = h_3 = h_4 = 0 \,, \notag
\end{align}
where $\xi(w)$ and $\zeta(w)$ are the leading-order IW functions for ground-state mesons and ground-state baryons, respectively, satisfying $\xi(1)=\zeta(1)=1$.
For the mesonic transition, we take corrections of the heavy quark expansion into account at $\mathcal{O}(\alpha_s,\,\bar \Lambda/m_{b,c},\,\bar \Lambda^2/m_{c}^2)$ \cite{Iguro:2020cpg}. 
For the baryonic transition, by following Refs.~\cite{Bernlochner:2018kxh, Bernlochner:2018bfn}, we take corrections of heavy quark expansion into account at $\mathcal{O}(\alpha_s,\,\bar \Lambda/m_{b,c},\,\alpha_s\bar \Lambda/m_{b,c},\,\bar \Lambda^2/m_{c}^2)$.

\section{Generic formula}
\label{sec:generic}

In this appendix, we provide a set of numerical formulae of the charged-lepton polarization and the forward-backward asymmetry.
They are defined by
\begin{align}
& P_l^{H_c} = \frac{\Gamma^{H_c}_{\lambda_l=1/2}-\Gamma^{H_c}_{\lambda_l=-1/2}}{\Gamma^{H_c}_{\lambda_l=1/2}+\Gamma^{H_c}_{\lambda_l=-1/2}},~~~~~
\Gamma^{H_c}_{\lambda_l} = \int_{1}^{w_{H_c,\,max}}\!\!dw\int_{-1}^{1}\! d\costl \frac{d^2\Gamma^{\lambda_l}(H_b \to H_cl\ov\nu)}{dw\,d\costl}, \\[0.5em]
& A_{\FB}^{H_cl} = \frac{\Gamma^{H_c}_{c_{\theta_l}>0}-\Gamma^{H_c}_{c_{\theta_l}<0}}{\Gamma^{H_c}_{c_{\theta_l}>0}+\Gamma^{H_c}_{c_{\theta_l}<0}},~~~~~
\Gamma^{H_c}_{c_{\theta_l} >(<) 0} = \sum_{\lambda_l = \pm 1/2} \int_{1}^{w_{H_c,\,max}}\!\!dw\int_{0\,(-1)}^{1\,(0)}\! d\costl \frac{d^2\Gamma^{\lambda_l}(H_b \to H_cl\ov\nu)}{dw\,d\costl},
\end{align}
with $w_{H_c,\,max} = (m_{H_b}^2+m_{H_c}^2-m_l^2)/(2m_{H_b}m_{H_c})$.
Then, the generic formulae of $(P_\tau^{H_c}/P_\tau^{H_c,\SM})(R_{H_c}/R_{H_c}^{\SM})$ and $(A_\FB^{H_c\tau}/A_\FB^{H_c\tau,\SM})(R_{H_c}/R_{H_c}^{\SM})$ are obtained as
\begin{align}
\label{Eq:Generic_first}
 \frac{P_\tau^{D}}{P_{\tau,\SM}^{D}} = 
  \left(\frac{R_D} {R_D^{\SM}}\right)^{-1} \!\!\times \Big(& |1+C_{V_L}^\tau|^2  + 3.02(0.01)|C_{S_L}^\tau+C_{S_R}^\tau|^2 + 0.16(0.01)|C_{T}^\tau|^2  \\
  &+ 4.47(0.02)\textrm{Re}[(1+C_{V_L}^\tau)(C_{S_L}^\tau+C_{S_R}^\tau)^*]  -1.06(0.03)\textrm{Re}[(1+C_{V_L}^\tau)C_{T}^{\tau*}] \Big)  \,, \nonumber\\[0.5em]
 \frac{P_\tau^{D^{\ast}}} {P_{\tau,\SM}^{D^{\ast}}} = 
  \left(\frac{R_{D^{\ast}}}  {R_{D^{\ast}}^{\SM}}\right)^{-1} \!\!\times \Big(& |1+C_{V_L}^\tau|^2   - 0.09(0.01)|C_{S_R}^\tau- C_{S_L}^\tau|^2 - 1.88(0.02) |C_{T}^\tau|^2 \\ 
   &+ 0.26(0.02) \textrm{Re}[(1+C_{V_L}^\tau)(C_{S_L}^\tau -C_{S_R}^\tau)^*] - 3.50(0.04) \textrm{Re}[(1+C_{V_L}^\tau) C_{T}^{\tau*}] \Big)\,, \nonumber \\[0.5em]
    \frac{P_\tau^{\Lambda_c}}{P_\tau^{\Lambda_c,\SM}}=\left(\frac{R_{\Lambda_c}}{R_{\Lambda_c}^{\SM}}\right)^{-1}\,\,\times \Big(& |1+C_{V_L}^\tau|^2
-1.48(0.01)\, \textrm{Re}[ (1 +C_{V_L}^\tau) C_{S_R}^{\tau*}]
-0.97(0.01) \,\textrm{Re}[ (1 +C_{V_L}^\tau) C_{S_L}^{\tau*}]\nonumber \\ 
&-1.47(0.02) \, \textrm{Re}[C_{S_L}^\tau C_{S_R}^{\tau*} ]
- 0.92(0.01) \, (|C_{S_L}^\tau|^2 + |C_{S_R}^\tau|^2) 
 \\[0.5em]
& -3.59(0.05) \,\textrm{Re}[ (1 +C_{V_L}^\tau)  C_T^{\tau*}]
-3.30(0.03)\, |C_T^\tau|^2\Big)  \,,\nonumber 
\end{align}
\begin{align}
   \frac{A_\FB^{D\tau}}{A_\FB^{D\tau,\SM}}=\left(\frac{R_{D}}{R_{D}^{\SM}}\right)^{-1}\,\,\times \Big(& |1+C_{V_L}^\tau|^2
+1.19(0.01)\, \textrm{Re}[ (1 +C_{V_L}^\tau) (C_{S_R}^\tau+C_{S_L}^\tau)^*] \nonumber \\ 
& +2.40(0.03) \,\textrm{Re}[ (1 +C_{V_L}^\tau)  C_T^{\tau*}]
 +3.04(0.04) \,\textrm{Re}[ (C_{S_R}^\tau+C_{S_L}^\tau)  C_T^{\tau*}]\Big)  \,,\\[0.5em]
\label{Eq:Generic_2ndlast}
\frac{A_\FB^{D^*\tau}}{A_\FB^{D^*\tau,\SM}}=\left(\frac{R_{D^*}}{R_{D^*}^{\SM}}\right)^{-1}\,\,\times \Big(& |1+C_{V_L}^\tau|^2
-2.61(0.60)\, \textrm{Re}[ (1 +C_{V_L}^\tau) (C_{S_R}^\tau-C_{S_L}^\tau)^*] +31.4(5.9) \,\textrm{Re}[ (1 +C_{V_L}^\tau)  C_T^{\tau*}]\nonumber \\ 
& -107(17) \,|C_{T}^\tau|^2
 +12.7(2.9) \,\textrm{Re}[ (C_{S_R}^\tau-C_{S_L}^\tau)  C_T^{\tau*}]\Big)  \,,\\[0.5em]
\label{Eq:Generic_last}
\frac{A_\FB^{\Lambda_c\tau}}{A_\FB^{\Lambda_c\tau,\SM}}=\left(\frac{R_{\Lambda_c}}{R_{\Lambda_c}^{\SM}}\right)^{-1}\,\,\times \Big(& |1+C_{V_L}^\tau|^2
+6.87(0.63)\, \textrm{Re}[ (1 +C_{V_L}^\tau) C_{S_R}^{\tau*}]
+1.87(0.12) \,\textrm{Re}[ (1 +C_{V_L}^\tau) C_{S_L}^{\tau*}] \nonumber\\[0.35em]
& -19.6(1.2) \,\textrm{Re}[ (1 +C_{V_L}^\tau)  C_T^{\tau*}]
+115(10)\, |C_T^\tau|^2  \\[0.35em]
& +26.1(2.6) \,\textrm{Re}[ C_{S_R}^\tau  C_T^{\tau*}]
+0.020(0.008)\,\textrm{Re}[ C_{S_L}^\tau  C_T^{\tau*}]  \Big)\,.\nonumber
\end{align}
The number in parentheses denotes the uncertainty, \eg, $3.02(0.01)$ means $3.02 \pm 0.01$, that comes from the form factor input and is approximated by Gaussian distributions.
The result agrees with Ref.\,\cite{Iguro:2024hyk} within uncertainty.

\section{More on the sum rule in the heavy quark limit}
\label{sec:alpha}

In the heavy quark limit, the sum rule for the double-differential decay rates of Eq.\,(\ref{eq:D3RSR}) can be transformed as
\begin{align}
 \label{eq:D3RSR_Ptau}
 &\frac{\kappa_{\Lambda_c}^{w,+1/2}-\kappa_{\Lambda_c}^{w,-1/2}}{\kappa_{\Lambda_c\,\SM}^{w,+1/2}-\kappa_{\Lambda_c\,\SM}^{w,-1/2}} = \alpha_{\rm HQL}^{P_\tau} \frac{\kappa_{D}^{w,+1/2}-\kappa_{D}^{w,-1/2}}{\kappa_{D\,\SM}^{w,+1/2}-\kappa_{D\,\SM}^{w,-1/2}} + \beta_{\rm HQL}^{P_\tau} \frac{\kappa_{D^*}^{w,+1/2}-\kappa_{D^*}^{w,-1/2}}{\kappa_{D^*\,\SM}^{w,+1/2}-\kappa_{D^*\,\SM}^{w,-1/2}}\,,\\[0.6em]
 \label{eq:D3RSR_AFB}
 &\frac{\kappa_{\Lambda_c+}^{w}-\kappa_{\Lambda_c-}^{w}}{\kappa_{\Lambda_c+}^{w\,\SM}-\kappa_{\Lambda_c-}^{w\,\SM}} = \alpha_{\rm HQL}^{A_\FB^\tau} \frac{\kappa_{D+}^{w}-\kappa_{D-}^{w}}{\kappa_{D+}^{w\,\SM}-\kappa_{D-}^{w\,\SM}}  + \beta_{\rm HQL}^{A_\FB^\tau} \frac{\kappa_{D^*+}^{w}-\kappa_{D^*-}^{w}}{\kappa_{D^*+}^{w\,\SM}-\kappa_{D^*-}^{w\,\SM}} \,,
\end{align}
where the differential decay rates in denominators are the SM ones, {\it i.e.}, $C_X^\tau = 0$. 
Also $\kappa_{H_c}^{w,\lambda_\tau}=\frac{d\Gamma^{H_c}_{\lambda_\tau}}{dw}$,  
$\kappa_{H_c\pm}^{w}=\pm \int^{\pm1}_{0} d\costa\frac{d^2\Gamma^{H_c}}{dwd\costa}$, and $\frac{d^2\Gamma^{H_c}}{dwd\costa}=\frac{d^2\Gamma^{H_c}_{+1/2}}{dwd\costa}+\frac{d^2\Gamma^{H_c}_{-1/2}}{dwd\costa}$ are defined.
The coefficients are given by
\begin{align}
 \label{eq:ab_HQL}
 \alpha_{\rm HQL}^{P_\tau} &= \frac{2}{1+w} \frac{\zeta(w)^2}{\xi(w)^2} \frac{\kappa_{D\,\SM}^{w,+1/2}-\kappa_{D\,\SM}^{w,-1/2}}{\kappa_{\Lambda_c\,\SM}^{w,+1/2}-\kappa_{\Lambda_c\,\SM}^{w,-1/2}}
 =\frac{(1 + r_D)^2 (1 + \rho^2_\tau   -w(1- 2 \rho^2_\tau)) - 6 r_D\rho^2_\tau (1 + w)}{4r_D(1-2\rho_\tau^2+w^2(2-\rho_\tau^2))-6w(1-\rho_\tau^2)(1+r_D^2)}  \,,\nonumber \\[0.6em]
 \alpha_{\rm HQL}^{A_\FB^\tau} &=
\frac{2}{1+w} \frac{\zeta(w)^2}{\xi(w)^2} \frac{\kappa_{D+}^{w\,\SM}-\kappa_{D-}^{w\,\SM}}{\kappa_{\Lambda_c+}^{w\,\SM}-\kappa_{\Lambda_c-}^{w\,\SM}} 
= \frac{-(1-r_D^2)\rho_\tau^2}{2(1-\rho_\tau^2+r_D^2(1+\rho_\tau^2)-2r_D w)}\,,
\end{align}
where $\rho_\tau={m_\tau}/\sqrt{q^2}$ and $r_i=m_i/m_B$ are defined.
They are independent of the leading-order IW functions $\zeta(w)$ and $\xi(w)$, and satisfy $\alpha_{\rm HQL}+\beta_{\rm HQL}=1$ as the sum rule should hold in the SM limit.
$\alpha_{\rm HQL}$ and $\beta_{\rm HQL}$ are independent of NP contributions $C_X$ and hence the sum rule holds in any NP model captured in Eq.\,(\ref{eq:Hamiltonian}).
Similar to the construction of the sum rule for $R_{H_c}$, the sum rules involving angular observables, Eqs.\,(\ref{eq:PtaSR}) and (\ref{eq:AFBSR}), are obtained by integrating numerators and denominators over $w$, and by normalizing with the decay widths of light-lepton modes.
It is noted that $\Gamma(H_b\to H_c \tau\ov\nu)$ dependence is canceled in the combinations of $P_\tau^{H_c}\,R_{H_c}$ and $A_\FB^{H_c}\,R_{H_c}$ for each.

\clearpage
\section{Figures}
\label{sec:plot}

In this appendix, we show the plot of probability distributions.
In each plot, we generated 20K Monte Carlo samples assuming the form factor parameters have Gaussian probability distributions. 
Figure \ref{fig:Pta} (Fig.\,\ref{fig:AFB}) shows the probability distribution of $\alpha_{P_\tau}$ ($\alpha_{A_\FB^\tau}$) and coefficients in the corrections to the sum rule, $\delta_{P_\tau}^{ij}$ ($\delta_{A_\FB^\tau}^{ij}$).
Here, the results are not normalized.
Also, correlations exist among the distributions, but they are not shown.
Figure \ref{fig:delPtaNP} (Fig.\,\ref{fig:delAFBNP}) shows the probability distribution of the correction $\delta_{P_\tau}$ ($\delta_{A_\FB^\tau}$) for the benchmark scenarios in Table\,\ref{Tab:result}.
The correction is zero in the $S_R$ and ${\rm{U}}_1$ LQ scenarios, and hence, is omitted.
\vspace{.5cm}
\begin{figure}[h]
\begin{center}
\includegraphics[width=0.31\linewidth]{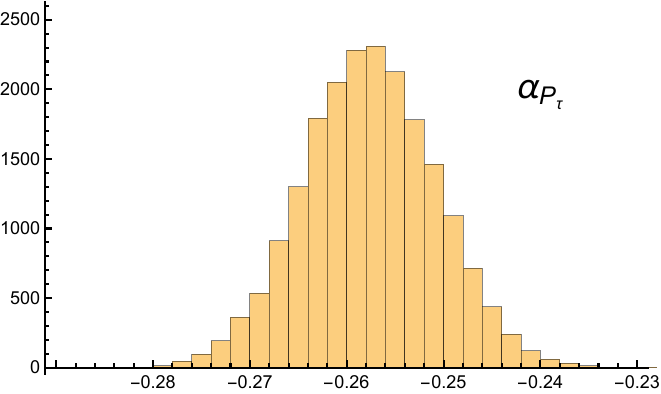}~
\includegraphics[width=0.31\linewidth]{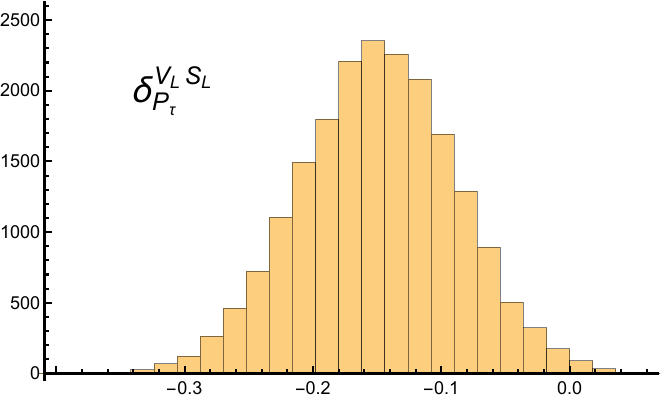}~
\includegraphics[width=0.31\linewidth]{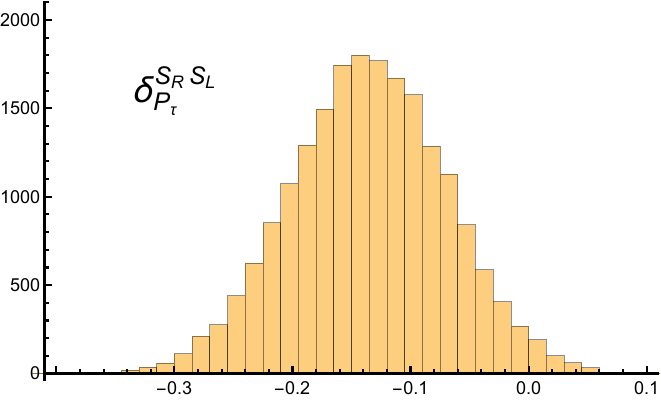}\\
\includegraphics[width=0.31\linewidth]{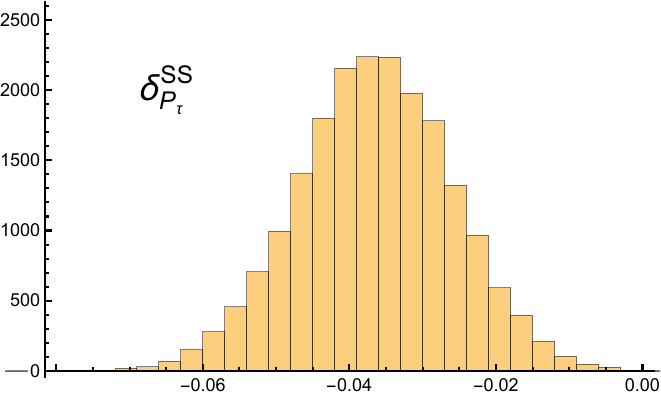}~
\includegraphics[width=0.31\linewidth]{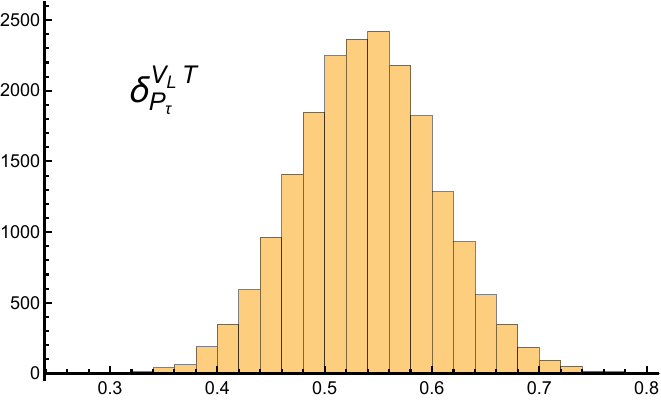}~
\includegraphics[width=0.31\linewidth]{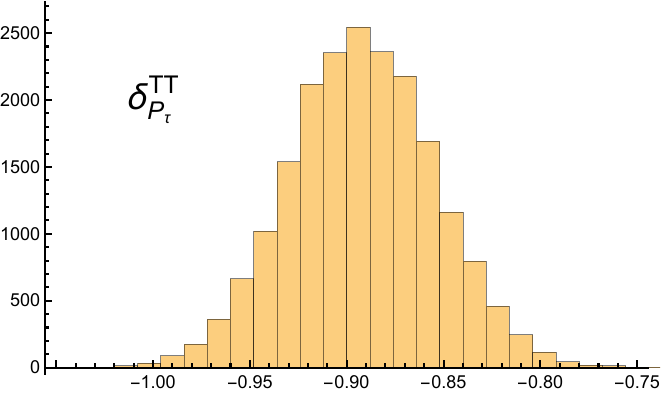}
\end{center}
\vspace{-.15cm}
\caption{
Probability distribution of $\alpha_{P_\tau}$ and the corrections $\delta_{P_\tau}^{ij}$. \\
}
\label{fig:Pta}
\end{figure}
\vspace{.5cm}

\begin{figure}[h]
\begin{center}
\includegraphics[width=0.31\linewidth]{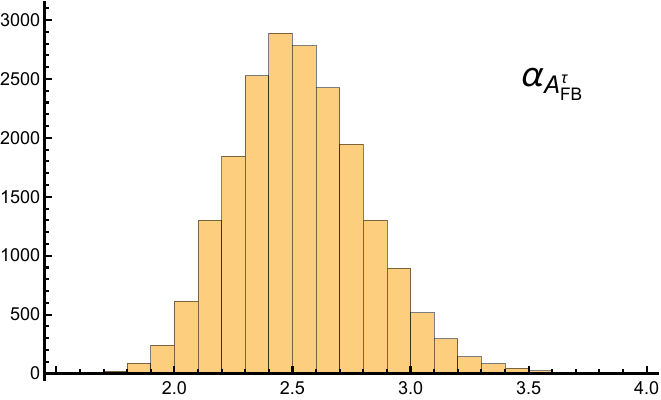}~
\includegraphics[width=0.31\linewidth]{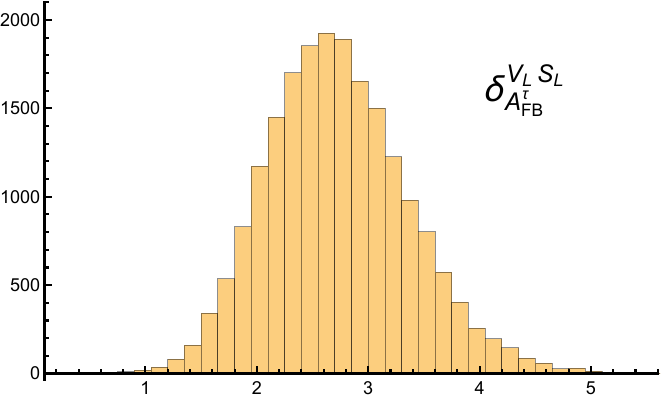}~
\includegraphics[width=0.31\linewidth]{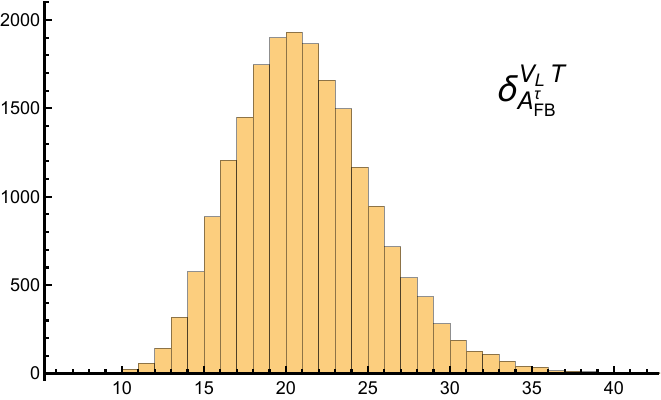}\\
\includegraphics[width=0.31\linewidth]{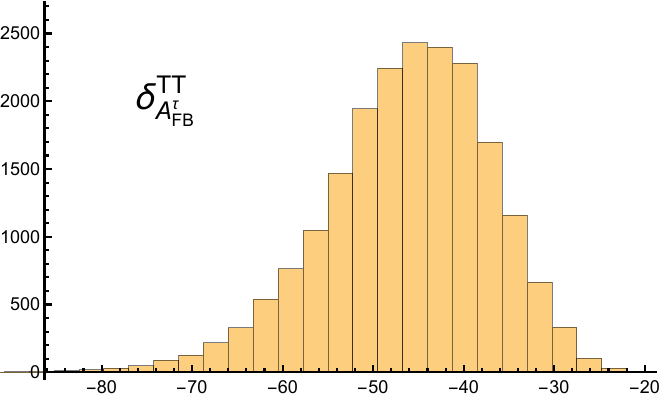}~
\includegraphics[width=0.31\linewidth]{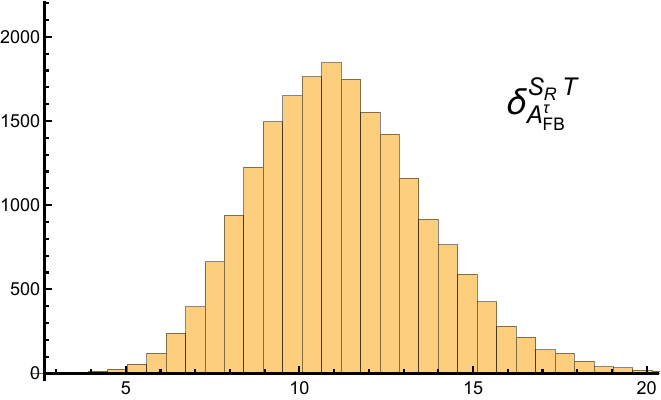}~
\includegraphics[width=0.31\linewidth]{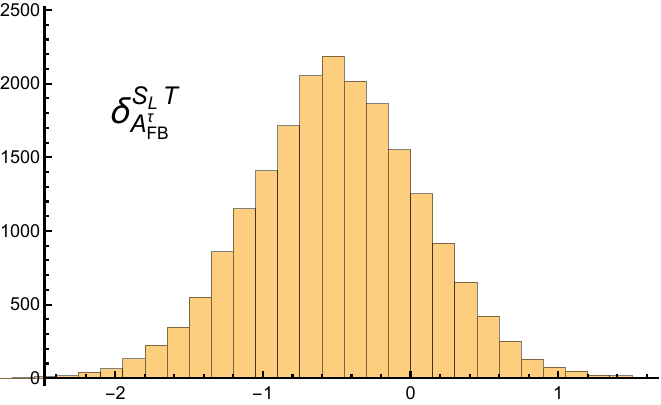}
\end{center}
\vspace{-.15cm}
\caption{
Probability distribution of $\alpha_{A_{\FB}^\tau}$ and the corrections $\delta_{A_{\FB}^\tau}^{ij}$.\\
}
\label{fig:AFB}
\end{figure}
\vspace{.5cm}

\begin{figure}[h]
\begin{center}
\includegraphics[width=0.31\linewidth]{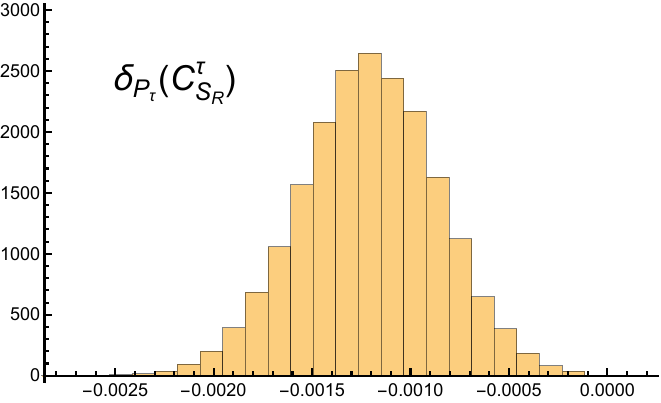}~
\includegraphics[width=0.31\linewidth]{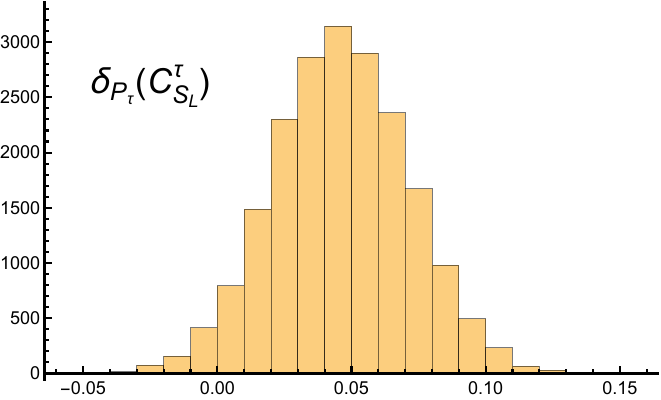}~
\includegraphics[width=0.31\linewidth]{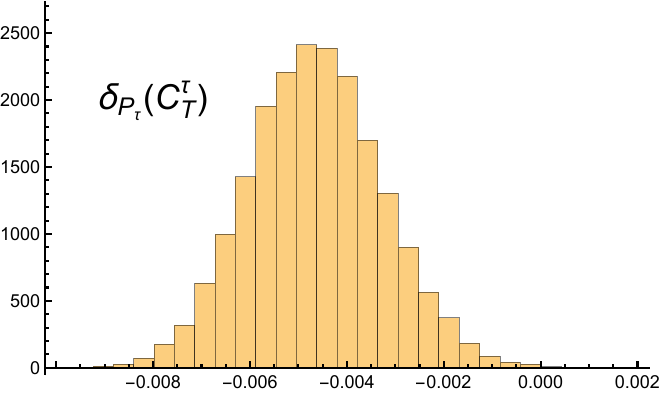}\\
\includegraphics[width=0.31\linewidth]{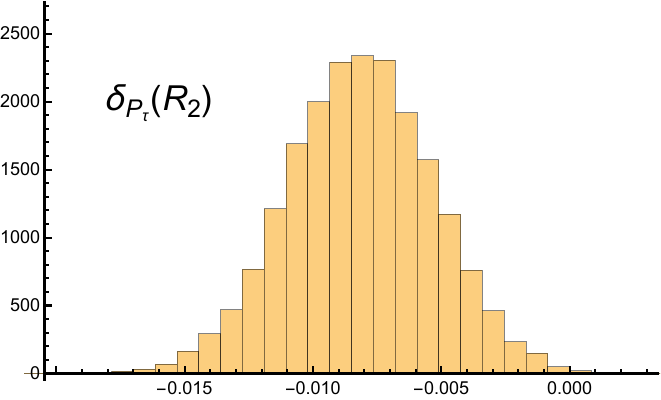}~
\includegraphics[width=0.31\linewidth]{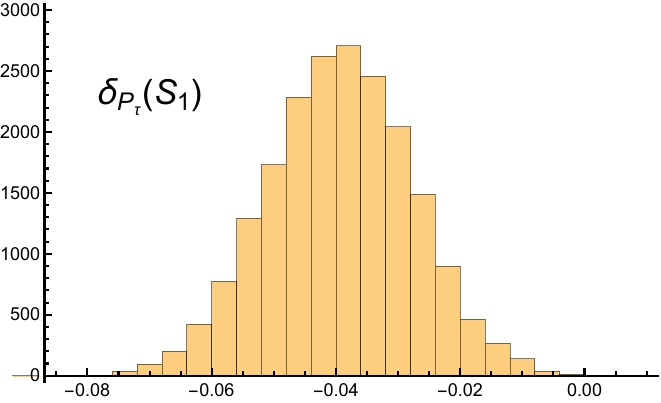}~
\includegraphics[width=0.31\linewidth]{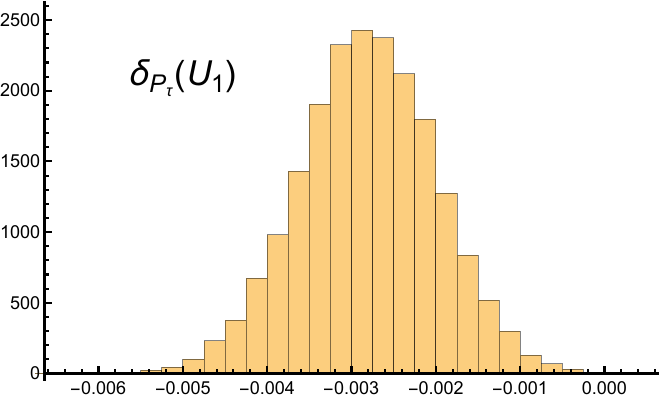}
\end{center}
\vspace{-.15cm}
\caption{
Probability distribution of the correction $\delta_{P_\tau}$ based on the benchmark NP scenarios shown in Table\,\ref{Tab:result}.\\
\vspace{.5cm}
}
\label{fig:delPtaNP}
\end{figure}
\vspace{.5cm}

\begin{figure}[h]
\begin{center}
\includegraphics[width=0.31\linewidth]{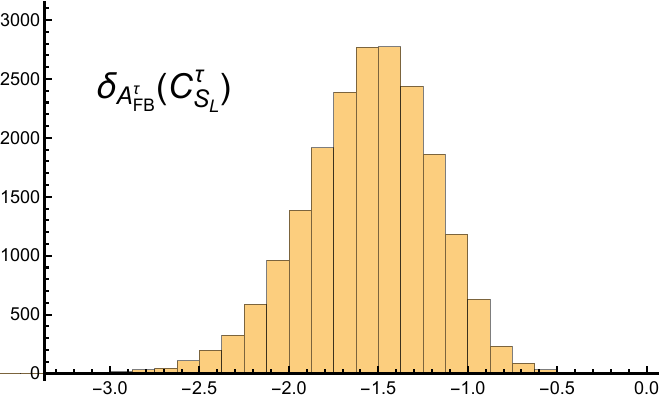}~
\includegraphics[width=0.31\linewidth]{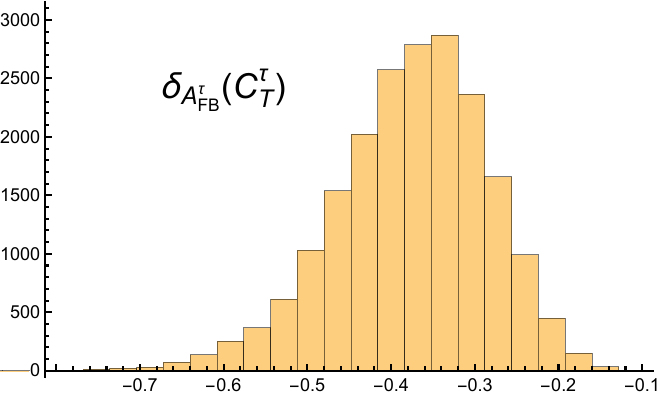}\\
\includegraphics[width=0.31\linewidth]{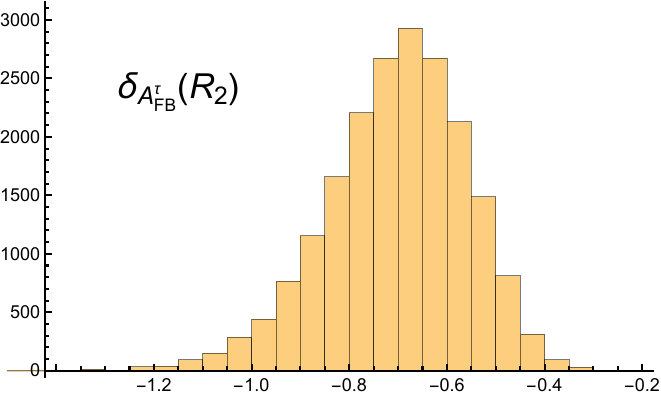}~
\includegraphics[width=0.31\linewidth]{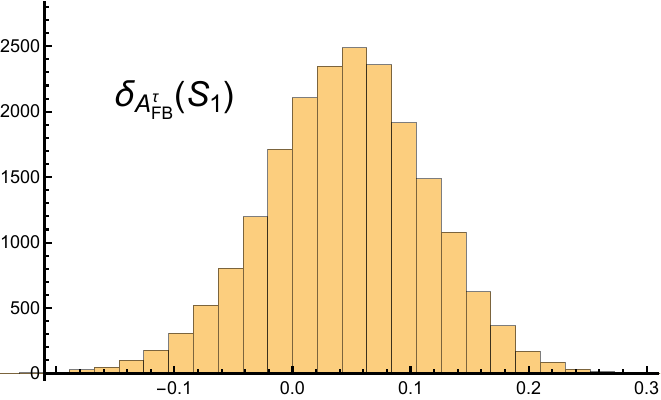}

\end{center}
\vspace{-.15cm}
\caption{
Probability distribution of the correction $\delta_{A_{\FB}^\tau}$ based on the benchmark NP scenarios shown in Table\,\ref{Tab:result}.\\
}
\label{fig:delAFBNP}
\end{figure}

\end{widetext}
\bibliographystyle{utphys}
\bibliography{ref}
\end{document}